\documentstyle[preprint,aps,floats]{revtex}
\tightenlines

\input epsf.tex
\def\DESepsf(#1 width #2){\epsfxsize=#2 \epsfbox{#1}}
\begin{document}
\preprint{\vbox{\hbox{}\hbox{KEK-TH-906} \hbox{hep-ph/0307334}}}
\draft
\title {Remarks on the recently observed $B$ decays \\
into $f_X(1300) K$ and $J/\psi K_X^*(1430)$ }

\author{C.~ S. ~Kim$^{1}$\footnote{cskim@yonsei.ac.kr}
and~~ Sechul Oh$^{2}$\footnote{scoh@post.kek.jp}}

\address{
$^1$Department of Physics and IPAP, Yonsei University, Seoul
120-479, Korea\\
$^2$Theory Group, KEK, Tsukuba, Ibaraki 305-0801, Japan}
\maketitle

\begin{abstract}
\noindent
In light of the recent experimental results on decays $B^+ \to f_X(1300) K^+$
and $B^{\pm(0)} \to J/\psi K_X^{*\pm(0)}(1430)$
from Belle, we study the
$B \to P(V) S$ type decays, $B^{\pm(0)} \to f_0(1370) K^{\pm(0)}$ and
$B^{\pm(0)} \to J/\psi K_0^{*\pm(0)}(1430)$,
in comparison with the $B \to P(V) T$ type
decays, $B^{\pm(0)} \to f_2(1270) K^{\pm(0)}$ and
$B^{\pm(0)} \to J/\psi K_2^{*\pm(0)}(1430)$.
We calculate the BRs for these decays by using the form factors obtained in
the ISGW2 model [the improved version of the original Isgur-Scora-Grinstein-Wise
(ISGW) model], as well as the ISGW model for comparison.
The ratios of ${\cal B}(B \to P(V)S) / {\cal B}(B \to P(V)T)$ are also presented.
\end {abstract}

%%%%%%%%%%%%%%%%%%%%%%%%%%%%%%%%%%%%%%%%%%%%%%%%%%%%%%%%%%
%%%%%%%%%%%%%%%%%%%%%%%%%%%%%%%%%%%%%%%%%%%%%%%%%%%%%%%%%%
\newpage
>From $B$ factory experiments such as Belle and BaBar, a tremendous amount of
experimental data on $B$ decays start to provide new bounds on previously known
observables with an unprecedented precision as well as an opportunity to see
very rare decay modes for the first time.
Experimentally several scalar mesons ($S$) have been observed \cite{Hagiwara:fs},
such as the isospinor $K_0^*$(1430), the isovectors $a_0$(980) and $a_0$(1450),
the isoscalars $\sigma$ or $f_0$(600), $f_0$(1370), $f_0$(1500), and heavier
scalar mesons.
Several tensor mesons ($T$) have been also observed \cite{Hagiwara:fs}: for
instance, the isovector $a_2$(1320), the isoscalars $f_2$(1270),
$f_2^{\prime}$(1525), $f_2$(2010), $f_2$(2300), $f_2$(2340),
$\chi_{c2}(1P)$, $\chi_{b2}(1P)$ and $\chi_{c2}(2P)$, and the
isospinors $K_2^*$(1430) and $D_2^*$(2460).
The measured branching ratios (BRs) for $B$ decays involving a pseudoscalar ($P$)
or a vector ($V$), and a tensor meson in the final state provide only upper
bounds \cite{Hagiwara:fs}.
In particular, the process $B \to K_2^* \gamma$ has been observed
for the first time by the CLEO Collaboration with a branching
ratio of $(1.66^{+0.59}_{-0.53} \pm 0.13) \times 10^{-5}$
\cite{Coan:1999kh}, and by the Belle Collaboration with
${\cal B}(B\to K_2^*\gamma)=(1.26\pm0.66\pm0.10)\times 10^{-5}$ \cite{belle0}.

Recently Belle reported the first observation of the decay $B^+ \to f_0(980) K^+$
\cite{Abe:2002av}, which is the first reported example of $B \to S P$ decay.
The BR of ${\cal B}(B^+\to K^+ \pi^- \pi^+)=(55.6\pm5.8\pm7.7)\times 10^{-6}$ was
also measured for the first time by Belle \cite{Abe:2002av}.  However, the
interpretation of possible states for a $\pi^+\pi^-$ invariant mass around 1300 MeV
in the $K^+ \pi^- \pi^+$ system remains unclear, even though two possible candidate
states have been suggested: $f_0(1370)$ and $f_2(1270)$.
The measured BR product for the $f_X(1300) K^+$ final state \cite{Abe:2002av} is
\begin{equation}
{\cal B}(B^+ \to f_X(1300) K^+) \times {\cal B}(f_X(1300) \to \pi^+ \pi^-)
= (11.1^{+3.4 +1.4 +7.2}_{-3.1 -1.4 -2.9}) \times 10^{-6} ~.
\label{fxKdata}
\end{equation}
At present, there exist some theoretical studies on the processes $B \to f_2(1270) K$
\cite{Kim:2002rx,Kim:2001sh}, but no theoretical information
on the processes $B \to f_0(1370) K$ exists.
Thus, it would be difficult to make clearer interpretation of the states for a
$\pi^+ \pi^-$ invariant mass around 1300 MeV in the $K^+ \pi^- \pi^+$ system.

Another type of nonleptonic $B$ decays involving a scalar or a tensor meson has
been very recently observed by Belle : $B^{\pm(0)} \to J/\psi K_X^{*\pm(0)}(1430)$,
where $K_X^*(1430)$ is $K_0^*(1430)$ or $K_2^*(1430)$ \cite{belle1}.
This is also the first observation of $B \to V S$ decay.
As the case of $B \to f_X(1300) K$, no theoretical information on the modes
$B^{\pm(0)} \to J/\psi K_0^{* \pm(0)}(1430)$ exists, while some theoretical works on
the modes $B^{\pm(0)} \to J/\psi K_2^{* \pm(0)}(1430)$ have been done
\cite{Kim:2002rw,Kim:2001py,LopezCastro:1997im}.
Therefore, the interpretation of the state $K_X^{*\pm(0)}(1430)$ would be in difficulty
and certain theoretical inputs would be necessary.  In particular, the ratio of
the BRs, ${\cal B}(B \to J/\psi K_0^*(1430)) / {\cal B}(B \to J/\psi K_2^*(1430))$,
would be very useful as such a theoretical input.

In light of the recent Belle results on $B^+ \to f_X(1300) K^+$ and
$B^{\pm(0)} \to J/\psi K_X^{*\pm(0)}(1430)$,
we study the $B \to P S$ and $B \to V S$ decays in
comparison with the corresponding $B \to P T$ and $B \to V T$ decays.
Our focus is on the particular processes
$B^{\pm(0)} \to f_0(1370) K^{\pm(0)}$ and $B^{\pm(0)} \to J/\psi K_0^{*\pm(0)}(1430)$
in comparison with the processes $B^{\pm(0)} \to f_2(1270) K^{\pm(0)}$ and
$B^{\pm(0)} \to J/\psi K_2^{*\pm(0)}(1430)$.

It is known \cite{scalar} that in contrast to the vector and tensor mesons, the
identification of the scalar mesons in experiment is difficult.  Also theoretically
the internal structure of most scalar mesons is not very clear
\cite{Godfrey:1998pd,Close:2002zu}.
Among light scalar mesons, $K_0^*(1430)$ is the least controversial and its
quark content is quite obvious.  In contrast, the quark content of $f_0(1370)$ is
relatively less clear: it is believed to be mainly ${1 \over \sqrt{2}}
(u \bar u + d \bar d)$, but the portion of its $s \bar s$ component is unknown
\cite{Cheng:2002mk,Cheng:2002ai}.
We set the quark content of $f_0(1370)$ as
\begin{eqnarray}
f_0(1370) = {1 \over \sqrt{2}} (u \bar u + d \bar d) \cos\phi_{_S} + s \bar s \sin\phi_{_S} ~,
\end{eqnarray}
where $\phi_{_S}$ denotes the mixing angle of the scalar meson $f_0(1370)$.
Similarly, the quark content of the tensor meson $f_2(1270)$ can be written as
\begin{eqnarray}
f_2(1270) = {1 \over \sqrt{2}} (u \bar u + d \bar d) \cos\phi_{_T} + s \bar s \sin\phi_{_T} ~,
\end{eqnarray}
where the mixing angle $\phi_{_T}$ is given by $\phi_{_T} =
\arctan(1 / \sqrt{2}) - 28^0 \approx 7^0$ \cite{Katoch:hr,Li:2000zb}.

The relevant $\Delta B =1$ effective Hamiltonian for hadronic $B$ decays
can be written as
\begin{eqnarray}
H_{eff}^{q} = {G_F \over \sqrt{2}} \left[ V_{ub}V^{*}_{uq} (c_1 O^q_{1u}
+c_2 O^q_{2u}) + V_{cb}V^{*}_{cq} (c_1 O^q_{1c} +c_2 O^q_{2c})
- V_{tb} V^*_{tq} \sum_{i=3}^{10} c_i O_i^q \right]  + H.c.~ ,
\end{eqnarray}
where $O^q_i$'s are defined as
\begin{eqnarray}
O^q_{1f} &=& \bar q \gamma_{\mu} L f \bar f \gamma^{\mu} L b,  \ \
O^q_{2f} = \bar q_{\alpha} \gamma_{\mu} L f_{\beta} \bar f_{\beta} \gamma^{\mu}
L b_{\alpha}~,   \nonumber \\
O^q_{3(5)} &=& \bar q \gamma_{\mu} L b \sum_{q^{\prime}} \bar q^{\prime}
\gamma^{\mu} L(R) q^{\prime},  \ \
O^q_{4(6)} = \bar  q_{\alpha} \gamma_{\mu} L b_{\beta} \sum_{q^{\prime}}
\bar q^{\prime}_{\beta}  \gamma^{\mu}
L(R) q^{\prime}_{\alpha}~,  \nonumber \\
O^q_{7(9)} &=& {3 \over 2} \bar q \gamma_{\mu} L b \sum_{q^{\prime}}
e_{q^{\prime}} \bar q^{\prime}  \gamma^{\mu}
R(L) q^{\prime} , \ \
O^q_{8(10)} ={3 \over 2} \bar q_{\alpha} \gamma_{\mu} L b_{\beta}
\sum_{q^{\prime}}  e_{q^{\prime}} \bar
q^{\prime}_{\beta} \gamma^{\mu} R(L) q^{\prime}_{\alpha}~ ,
\end{eqnarray}
where $L(R) = (1 \mp \gamma_5)$, $f$ can be $u$ or $c$  quark, $q$
can be $d$ or $s$ quark,  and $q^{\prime}$ is summed over $u$,
$d$, $s$, and $c$ quarks. $\alpha$ and  $\beta$ are  the color
indices.  $T^a$ is the SU(3) generator with the normalization
${\rm Tr}(T^a T^b) =  \delta^{ab}/2$. $G^{\mu \nu}_a$ and $F^{\mu
\nu}$ are the gluon and photon field strength, and $c_i$'s  are
the Wilson  coefficients (WCs). We use the improved effective
WCs given in Ref.\cite{Ali:1997nh,Chen:1999nx}, where the renormalization
scheme- and scale-dependence of the WCs are discussed and
resolved.  The regularization scale is taken to be $\mu=m_b$
\cite{Deshpande:1997ar}. The operators $O_1$, $O_2$ are the tree level and QCD
corrected operators, $O_{3-6}$ are the gluon induced strong
penguin operators, and finally  $O_{7-10}$ are the electroweak
penguin operators due to $\gamma$ and $Z$ exchange, and  the box
diagrams at loop level.

To calculate the BRs of the interested decay processes, we adopt the
generalized factorization framework.   The hadronic matrix elements for
$B \to P(V)S$ and $B \to P(V)T$ decays can be parameterized as
\begin{eqnarray}
\langle 0 | A^{\mu} | P \rangle &=& i f_P p_P^{\mu} ~, ~~~~~
\langle 0 | V^{\mu} | S \rangle =  f_S p_S^{\mu} ~, \\
\langle 0 | V^{\mu} | V \rangle &=& f_{_V} m_{_V} \epsilon^{\mu} ~,  \\
\langle S | A^{\mu} | B \rangle &=& F_+^{B \to S}(q^2) (p_B + p_S)^{\mu}
 + F_-^{B \to S}(q^2) (p_B - p_S)^{\mu} ~,
 \label{BtoSformfactor}  \\
\langle T | j^{\mu} | B \rangle &=& i h(q^2) \epsilon^{\mu \nu
 \rho \sigma} \epsilon^*_{\nu \alpha} p_B^{\alpha} (p_B
 +p_T)_{\rho} (p_B -p_T)_{\sigma} + k(q^2) \epsilon^{* \mu \nu}
 (p_B)_{\nu}  \nonumber \\
 &\mbox{}&  + \epsilon^*_{\alpha \beta} p_B^{\alpha} p_B^{\beta} [
 b_+(q^2) (p_B +p_T)^{\mu} +b_-(q^2) (p_B -p_T)^{\mu} ]~,
\label{formfactor}
\end{eqnarray}
where $j^{\mu} = V^{\mu} -A^{\mu}$.  $V^{\mu}$ and $A^{\mu}$
denote a vector and an axial-vector current, respectively.
The $f_M ~(M = P, ~S, ~V)$ denotes the decay constant of the relevant meson $M$.
The $p_M^{\mu} ~(M = B, ~P, ~S, ~T)$ denotes the four-momentum of the relevant meson
$M$.
Here $\epsilon^\mu (\epsilon^{\mu\nu})$ is the polarization vector (tensor) of
the vector (tensor) meson.
The $F_{\pm}^{B \to S}(q^2)$ are the form factors for the $B \to S$ transition,
and $h(q^2)$, $k(q^2)$, and $b_{\pm}(q^2)$ are the form factors
for the $B \to T$ transition, which are calculated at $q^2$
$(q^{\mu} \equiv p_B^{\mu} -p_{S(T)}^{\mu})$.
The form factors $F_{\pm}^{B \to S}$, $h$, $k$, $b_\pm$ contain nonperturbative
nature of the $B \to S$ and $B \to T$ transitions, and in general they are functions
of the momentum transfer $q^2 \equiv (p_B-p_{S(T)})^2$.

Note that $\langle 0|V^{\mu}|f_0 \rangle = 0$,
since the decay constants of neutral scalars must vanish (i.e., $f_{f_0} =0$)
owing to charge conjugation invariance or conservation of vector current
\cite{Diehl:2001xe}.
The decay constant of $K_0^{*+}$ is not zero, but suppressed: for example, from
the finite-energy sum rules \cite{Maltman:1999jn}, $f_{K_0^{*+}} = 42$ MeV.
For tensor mesons, the following relation holds:
\begin{equation}
\langle 0|j^\mu|T \rangle = p_\nu\epsilon^{\mu\nu}(p_T,\lambda)
 +p_T^\mu\epsilon^\nu_\nu(p_T,\lambda)=0~.
\end{equation}
Thus, there are no amplitudes proportional to $f_{T}(f_{f_0}) \times
[{\rm form~factor~for~} B \to P(V)]$.
Using the above parameterizations, the decay amplitudes for $B \to P(V) f_0$ and
$B \to P(V) T$ \cite{Kim:2002rw,Oh:1998wa} are
\begin{eqnarray}
&\mbox{}& {\cal A}(B \to P f_0) \sim F_+^{B \to f_0}(q^2)~,~~~~~
{\cal A}(B \to V f_0) \sim (\epsilon_{\mu} p_B^{\mu}) F_+^{B \to f_0}(q^2)~,
\nonumber \\
&\mbox{}& {\cal A}(B \to P T) \sim {\cal F}^{B \to T}(q^2)~,~~~~~
{\cal A}(B \to V T) \sim \epsilon^{*\alpha\beta} {\cal F}^{B \to T}_{\alpha\beta}
(q^2)~,
\label{ampformfactor}
\end{eqnarray}
where
\begin{eqnarray}
{\cal F}^{B \to T}(q^2) &=& k(q^2)+(m_B^2-m_T^2) b_+(q^2)+ m_P^2 b_-(q^2)~,
\label{FBT} \\
{\cal F}^{B \to T}_{\alpha\beta}(q^2) &=& \epsilon^*_\mu( p_B+p_T)_\rho
 \Big[ih(q^2)\cdot \epsilon^{\mu\nu\rho\sigma}
  g_{\alpha\nu}(p_V)_\beta(p_V)_\sigma
 + k(q^2)\cdot\delta^\mu_\alpha\delta^\rho_\beta\nonumber\\
&& +b_+(q^2) \cdot(p_V)_\alpha(p_V)_\beta g^{\mu\rho}\Big]~.
\label{FBTv}
\end{eqnarray}
The definition of $F_+^{B \to f_0}$ is given in Eq. (\ref{BtoSformfactor}).
In passing, we note that in factorization the decay amplitude of $B \to J/\psi K_0^*$
has no term proportional to $f_{K_0^*}$, so it has the same structure as
${\cal A}(B \to V f_0)$:
i.e., ${\cal A}(B \to J/\psi K_0^*) \sim (\epsilon_{\mu} p_B^{\mu})
{\cal F}_+^{B \to K_0^*}(q^2)$, where $\epsilon_{\mu}$ is the polarization
vector of $J/\psi$.

As seen in Eq. (\ref{ampformfactor}), the decay amplitudes (and subsequently
the BRs) are heavily dependent on the hadronic form factors which are model-dependent.
To compute the $B \to S$ and $B \to T$ form factors, we use the ISGW2 model
\cite{Scora:1995ty} which is the improved version of the nonrelativistic quark
model of Isgur, Scora, Grinstein and Wise (ISGW) \cite{Isgur:gb}.
For comparison, we also compute the form factors using the original ISGW model.
A characteristic feature of the form factors given in the
original ISGW model is that values of the form factors decrease exponentially
as a function of $(q_m^2 - q^2)$, where $q^2 \equiv (p_B -p_{S(T)})^2$ is the
momentum transfer and $q_m^2 \equiv (m_B -m_{S(T)})^2$ is the maximum possible
momentum transfer in the $B$ meson rest frame for a $B \to S(T)$ transition.
This feature leads to the unreasonably small form factors at $q^2 = 0$, so
the form factors are sometimes calculated at the maximum momentum transfer
$q_m^2$, assuming that in the relevant transitions the momentum transfer
$(q^2)$ is close to the maximum momentum transfer $(q_m^2)$.
The ISGW model has been improved to the ISGW2 model in which the form factors
have a more realistic behavior at large $(q_m^2 -q^2)$ by making the replacement
of the exponentially decreasing term to a certain polynomial term
\cite{Scora:1995ty}.

Now we consider the decay processes $B^{\pm(0)} \to J/\psi K_0^{* \pm(0)}(1430)$
and $B^{\pm(0)} \to f_0(1370) K^{\pm(0)}$ in comparison with
$B^{\pm(0)} \to J/\psi K_2^{* \pm(0)}(1430)$ and $B^{\pm(0)} \to f_2(1270)
K^{\pm(0)}$.  The relevant decay amplitudes are given by
\begin{eqnarray}
A(B^{+(0)} \to J/\psi K_0^{*+(0)}(1430))
 &=& -i {G_F \over \sqrt{2}} 2 m_{J/\psi} f_{J/\psi} (\epsilon^{*} \cdot p_B)
 F_+^{B \to K_0^{*+(0)}}
\label{jpsik0} \nonumber \\
 &\mbox{}& \times [ V^*_{cb} V_{cs} a_2
 - V^*_{tb} V_{ts} (a_3 +a_5 +a_7 +a_9 ) ], \\
A(B^{+(0)} \to J/\psi K_2^{*+(0)}(1430))
 &=& -i {G_F \over \sqrt{2}} m_{J/\psi} f_{J/\psi} \epsilon^{* \alpha \beta}
 {\cal F}_{\alpha \beta}^{B \to K_2^{*+(0)}}
\label{jpsik2} \nonumber   \\
 &\mbox{}& \times [ V^*_{cb} V_{cs} a_2
 - V^*_{tb} V_{ts} (a_3 +a_5 +a_7 +a_9 ) ], \\
A(B^+ \to f_0(1370) K^+)
 &=& -i {G_F \over 2} \cos\phi_{_S} (m_B^2 - m_{f_0}^2) f_K F_+^{B \to f_0}
\label{f0kplus}  \nonumber \\
 &\mbox{}& \times \left\{ V^*_{ub} V_{us} a_1
 - V^*_{tb} V_{ts} [ a_4 +a_{10} - 2 (a_6 +a_8) X_{su} ] \right\}, \\
A(B^0 \to f_0(1370) K^0)
 &=& i {G_F \over 2} \cos\phi_{_S} (m_B^2 - m_{f_0}^2) f_K F_+^{B \to f_0}
\label{f0kzero}  \nonumber \\
 &\mbox{}& \times \left\{ V^*_{tb} V_{ts} \left[ a_4 -{1 \over 2} a_{10}
 - 2 \left( a_6 -{1 \over 2} a_8 \right) X_{sd} \right] \right\}, \\
A(B^+ \to f_2(1270) K^+)
 &=& -i {G_F \over 2} \cos\phi_{_T} (\epsilon^*_{\mu \nu} p_B^{\mu} p_B^{\nu})
 f_K {\cal F}^{B \to f_2}
\label{f2kplus}  \nonumber \\
 &\mbox{}& \times \left\{ V^*_{ub} V_{us} a_1
 - V^*_{tb} V_{ts} [ a_4 +a_{10} - 2 (a_6 +a_8) X_{su} ] \right\}, \\
A(B^0 \to f_2(1270) K^0)
 &=& i {G_F \over 2} \cos\phi_{_T} (\epsilon^*_{\mu \nu} p_B^{\mu} p_B^{\nu})
 f_K {\cal F}^{B \to f_2}   \nonumber \\
 &\mbox{}& \times \left\{ V^*_{tb} V_{ts} \left[ a_4 -{1 \over 2} a_{10}
 - 2 \left( a_6 -{1 \over 2} a_8 \right) X_{sd} \right] \right\},
\label{f2kzero}
\label{f2k0}
\end{eqnarray}
where
\begin{eqnarray}
X_{sq} = {m_K^2 \over (m_b +m_q)(m_s +m_q) } ~~~~~~ (q = u, ~d)~.
\end{eqnarray}
Here the effective coefficients $a_i$ are defined as $a_i =
c^{eff}_i + \xi c^{eff}_{i+1}$ ($i =$ odd) and $a_i = c^{eff}_i +
\xi c^{eff}_{i-1}$ ($i =$ even) with the effective WC's
$c^{eff}_i$ at the scale $m_b$ \cite{Deshpande:1997ar,Dutta:1999tg},
and by treating $\xi\equiv 1/N_c$ ($N_c$ denotes the effective number of colors)
as an adjustable parameter.  The terms with $F_-^{B \to S}$ and $b_-$ are
neglected because they give negligible contributions to the decay amplitudes
due to the small mass factor.  We have assumed that in $B \to f_{0(2)}K$
the weak annihilation contribution can be neglected compared to the tree
contribution.

{}From Eqs. (\ref{jpsik0}) to (\ref{f2k0}), it is obvious that the ratios of
the BRs, ${\cal B}(B \to J/\psi K_0^{*}(1430)) /
{\cal B}(B \to J/\psi K_2^{*}(1430))$ and
${\cal B}(B \to f_0(1370) K) / {\cal B}(B \to f_2(1270) K)$, are independent of
the parameter $\xi$, though they are still sensitive to the form factors:
\begin{eqnarray}
{ {\cal B}(B^{+(0)} \to J/\psi K_0^{*+(0)}(1430)) \over
  {\cal B}(B^{+(0)} \to J/\psi K_2^{*+(0)}(1430)) }
 &=& { \left| 2 (\epsilon^{*} \cdot p_B) F_+^{B \to K_0^{*+(0)} } \right|^2 \over
  \left| \epsilon^{* \alpha \beta} {\cal F}_{\alpha \beta}^{B \to K_2^{*+(0)}}
  \right|^2 }~,
  \nonumber \\
{\rm and}~~~~~~~{ {\cal B}(B^{+(0)} \to f_0(1370) K^{+(0)}) \over
  {\cal B}(B^{+(0)} \to f_2(1270) K^{+(0)}) }
 &=& { \left| \cos\phi_{_S} (m_B^2 - m_{f_0}^2) F_+^{B \to f_0} \right|^2 \over
  \left| \cos\phi_{_T} (\epsilon^*_{\mu \nu} p_B^{\mu} p_B^{\nu})
  {\cal F}^{B \to f_2} \right|^2 }~.
\end{eqnarray}

First, let us discuss magnitudes of the form factors in both models, the ISGW
and its improved version ISGW2.
In Table I, we show the values of the form factors $F_+^{B \to K_0^* (f_0)}$
and ${\cal F}^{B \to f_2}$ calculated in three cases:
(i) at $q^2 = m_{J/\psi}^2$ or $m_K^2$ $(q^{\mu} \equiv p_B^{\mu} -p_{S(T)}^{\mu})$
in the ISGW model,
(ii) at the maximum momentum transfer $q_m^2 \equiv (m_B -m_{S(T)})^2$
in the ISGW model, and
(iii) at $q^2 = m_{J/\psi}^2$ or $m_K^2$ in the ISGW2 model.
We note that in the ISGW model, $|{\cal F}^{B \to f_2(1270)}| =0.19$ at $q_m^2$,
while $|{\cal F}^{B \to f_2(1270)}| =0.025$ at $q^2 = m_K^2$.
The value of $|{\cal F}^{B \to f_2(1270)}|$ calculated at $q_m^2$ is 7.6 times
larger than that calculated at $q^2 = m_K^2$. Thus, the BR of a relevant
process (e.g., $B \to f_2(1270) K$) evaluated by using the former value of the
form factor (evaluated at $q_m^2$) would be roughly 60 times
larger than that obtained by using the latter value of the form factor
(at $q^2 = m_K^2$).  In contrast, in the ISGW2 model, $|{\cal F}^{B \to f_2(1270)}|
=0.078$ at $q^2 = m_K^2$ whose magnitude is in between that calculated at $m_K^2$
and that evaluated at $q_m^2$ in the ISGW model.
This feature is quite common in $B \to PT$ and $B \to VT$ decays (See Table I of
Ref. \cite{Kim:2002rw}).  It is because as previously mentioned, a crucial
improvement of the ISGW2 model is that the form factors in this model have a more
realistic behavior at large $(q_m^2 -q^2)$, by changing the exponential factor
of the form factors into a polynomial.
Subsequently, the BR calculated in the ISGW2 model is usually in between that
obtained at $q^2 =m_{P(V)}^2$ and that evaluated at zero recoil $q_m^2$ in the
ISGW model (Table III).

However, we find that for the $B \to S$ form factors, the situation is somewhat
different.  From Table I, we see that
\begin{eqnarray}
F_+^{B \to S} ({\rm in ~ ISGW2}) ~
 \lesssim ~ F_+^{B \to S} ({\rm at}~ q^2 =m_{P(V)}^2 ~ {\rm in ~ ISGW}) ~
 < ~ F_+^{B \to S} ({\rm at}~ q_m^2 ~ {\rm in ~ ISGW}) ,
\end{eqnarray}
for $S = K_0^{*+(0)}(1430), ~ f_0(1370)$.
Notice that the values of the form factors calculated in ISGW2 are
similar to or even smaller than those obtained at $q^2 =m_{P(V)}^2$ in ISGW.
This is not the case for the $B \to T$ form factors.
The main reason why it happens for the $B \to S$ form factors is that in the
ISGW2 model, there is an internal cancellation between two relevant terms in
$F_+^{B \to S}$ \cite{Scora:1995ty}, while in the ISGW model, no such cancellation
appears \cite{Isgur:gb}. Thus, in spite of its moderate behavior at large
$(q_m^2 -q^2)$ in ISGW2, the form factor $F_+^{B \to S} ({\rm in ~ ISGW2})$
becomes similar to or even smaller than
$F_+^{B \to S}({\rm at}~ q^2 =m_{P(V)}^2 ~{\rm in ~ ISGW})$.
Consequently, the relevant BRs computed in ISGW2 are similar to or even smaller
than those computed at $q^2 =m_{P(V)}^2$ in ISGW, and much smaller than those
obtained at the maximum momentum transfer $q_m^2$ in ISGW (Table II and III).
There is one more comment on $F_+^{B \to S}$ shown in Table I:
for $B \to f_0(1370) K$, $(q_m^2 - q^2) = 15.0$ GeV$^2$, while for
$B \to J/\psi K_0^*(1430)$, $(q_m^2 - q^2) = 5.4$ GeV$^2$.  Subsequently,
in the ISGW model, the difference in $F_+^{B \to f_0(1370)}$ between two cases,
at $q^2 =m_K^2$ and at $q_m^2$, is much larger than the corresponding difference
in $F_+^{B \to K_0^*(1430)}$.

Table II shows the BRs of $B \to J/\psi K_0^*(1430)$ and
$B \to J/\psi K_2^*(1430)$, computed in the ISGW2 model.  For comparison,
the BRs computed in the ISGW model (using $a_2 =0.26$) are also shown.
In the table, the results are shown for three different values of the parameter
$\xi$ : $\xi =0.1,~ 0.3,~ 0.5$~.  For comparison, the BRs are also calculated
for $a_2 \equiv c_2^{eff} + \xi c_1^{eff} =0.26$ whose values are obtained
from a fit to $B \to PP$ and $B \to PV$ data \cite{Alam:bi}.
The value of $a_2 =0.26$ corresponds to $\xi = 0.54$.
These decay modes are (color-suppressed) tree-dominant processes and their
decay amplitudes are dominantly proportional to the effective coefficients
$a_2$, as shown in Eqs. (\ref{jpsik0}) and (\ref{jpsik2}).
Since the value of $a_2$ becomes very small for $\xi =0.3$ due to a large
cancellation between $c_2^{eff}$ and $\xi c_1^{eff}$, the BRs for
$\xi =0.3$ are much smaller than those for $\xi =0.1$ or $\xi =0.5$.
>From the table, we see that the BRs strongly depend on the relevant form
factors.  Using $a_2 =0.26$, the BRs of $B^{+(0)} \to J/\psi K_0^{*+(0)}$
are about $7 \times 10^{-8}$ in the ISGW2 model, about $13 \times 10^{-8}$
at $q^2 = m_{J/\psi}^2$ and about $34 \times 10^{-8}$ at $q_m^2$ in the
ISGW model.  In contrast, the BRs of $B^{+(0)} \to J/\psi K_2^{*+(0)}$
are at least an order of magnitude larger than those of the corresponding
$B^{+(0)} \to J/\psi K_0^{*+(0)}$ modes: using $a_2 =0.26$,
${\cal B}(B^{+(0)} \to J/\psi K_2^{*+(0)}) = (1 - 4) \times 10^{-6}$.
Therefore, it is expected that the ratios of
${\cal B}(J/\psi K_0^{*+(0)}(1430)) / {\cal B}(J/\psi K_2^{*+(0)}(1430))$
are very small: about $(2 - 10) \%$.
These uncertainties arise only from the model-dependent form factors.

The BRs of $B^{+(0)} \to f_0(1370) K^{+(0)}$ and $B^{+(0)} \to f_2(1270)
K^{+(0)}$ are presented in Table III.  Unlike $B \to J/\psi K_{0(2)}^*$
decays, these decays $B \to f_{0(2)} K$ are penguin-dominant processes.
The charged modes $B^+ \to f_0(1370) K^+$ and $f_2(1270) K^+$ have
tree contributions proportional to $a_1$ as well as dominant penguin
contributions, while the neutral modes
$B^0 \to f_0(1370) K^0$ and $f_2(1270) K^0$ are pure penguin processes,
as shown in Eqs. (\ref{f0kplus})$-$(\ref{f2kzero}).
Because there appears a large cancellation between $a_4$ and $a_6$ in the
penguin amplitudes of $B^{+(0)} \to f_0(1370) K^{+(0)}$ and $f_2(1270)
K^{+(0)}$, the BRs of these modes become relatively small:
$O(10^{-8})-O(10^{-10})$ for $B^{+(0)} \to f_0(1370) K^{+(0)}$ and
$O(10^{-7})-O(10^{-9})$ for $B^{+(0)} \to f_2(1270) K^{+(0)}$.
In particular, the BRs of the neutral modes are much smaller because these
modes have only penguin contributions.
In Table III, we have used $\cos \phi_{_s} \approx 1$, which is a reasonable
approximation, because the scalar meson $f_0(1370)$ is believed to be
mainly composed of $u \bar u$ and $d \bar d$.
In fact, even in case of assuming some sizable portion of the $s \bar s$
component of $f_0(1370)$, the BRs of $B \to f_0(1370) K$ do not change much:
e.g., for $\phi_{_s}=18^0$ \cite{Cheng:2002mk}, these BRs change by only
a few percent.
The ratio of ${\cal B}(f_0(1370) K^{+(0)}) / {\cal B}(f_2(1270) K^{+(0)})$
is independent of $\xi$ and shown to be 0.16 in the ISGW2 model.
(But, in ISGW, the ratio is larger than 1.)
The CP rate asymmetries $A_{CP}$ for $B \to f_{0(2)} K$ are shown to be
very small in most cases: 0\%$-$3\%.

We note that for $B \to f_0(1370) K$ and $B \to f_2(1270) K$ decays, our
prediction given in Table III is strongly model-dependent.  Compared with
the Belle data shown in Eq. (\ref{fxKdata}), the BRs for $B \to f_0(1370) K$
decays predicted in the ISGW2 model seem to be very small.
In particular, the BRs for $B \to f_{0(2)} K$ calculated at the maximum
momentum transfer $q_m^2$ in ISGW model become about 40 (6) times larger
than those calculated at $q^2 = m_K^2$ in the ISGW2 model.  In this case
(i.e., for $q_m^2$ in the ISGW model), both BRs for $B^+ \to f_0(1370) K^+$
and for $B^+ \to f_2(1270) K^+$ are an order of $10^{-6}$, which would be
closer to the Belle data value.  However, we would need more caution to
seriously take these predicted values.  Clearly more reliable values for
the relevant form factors are called for from future studies.

To conclude, we have studied the $B \to P(V) S$ type decays,
$f_0(1370) K$ and $B \to J/\psi K_0^*(1430)$, in comparison with the
$B \to P(V) T$ type decays, $f_2(1270) K$ and $B \to J/\psi K_2^*(1430)$.
To calculate the relevant hadronic form factors, we have used the ISGW2
model as well as the original ISGW model for comparison.
The estimated BRs of these decays are sensitive to the model-dependent
form factors.  Using the ISGW2 model, the BRs are found to be
${\cal B}(B^{+(0)} \to J/\psi K_0^{*+(0)}(1430)) \approx 7 \times 10^{-6}$
for $a_2 =0.26$ and ${\cal B}(B^+(B^0) \to f_0(1370) K^+(K^0)) = 3.58(0.23)
\times 10^{-8}$ for $\xi =0.5$, while ${\cal B}(B^{+(0)} \to J/\psi
K_2^{*+(0)}(1430)) \approx 400 \times 10^{-6}$ for $a_2 =0.26$
and ${\cal B}(B^+(B^0) \to f_2(1270) K^+(K^0)) = 21.85(1.38) \times 10^{-8}$
for $\xi =0.5$.
The ratios of ${\cal B}(P(V)S) / {\cal B}(P(V)T)$ are independent of $\xi$,
but model-dependent: in ISGW2,
${\cal B}(J/\psi K_0^{*+(0)}(1430)) / {\cal B}(J/\psi K_2^{*+(0)}(1430))
\approx 2 \%$ and ${\cal B}(f_0(1370) K^{+(0)}) / {\cal B}(f_2(1270)
K^{+(0)}) = 16 \%$. \\

%%%%%%%%%%%%%%%%%%%%%%%%%%%%%%%%%%%%%%%%%%%%%%%%%%%%%%
%%%%%%%%%%%%%%%%%%%%%%%%%%%%%%%%%%%%%%%%%%%%%%%%%%%%%%
\centerline{\bf ACKNOWLEDGEMENTS}
\medskip

\noindent
We thank J. B. Singh and N. Soni for valuable discussions.
The work of C.S.K. was supported
by Grant No. R02-2003-000-10050-0 from BRP of the KOSEF.
The work of S.O. was supported by the Japan Society for the Promotion of
Science (JSPS).

%%%%%%%%%%%%%%%%%%%%%%%%%%%%%%%%%%%%%%%%%%%%%%%%%%%%%%
%%%%%%%%%%%%%%%%%%%%%%%%%%%%%%%%%%%%%%%%%%%%%%%%%%%%%%

%%%%%%%%%%%%%%%%%%%%%%%    Table 1    %%%%%%%%%%%%%%%%%%%%%%%%%%%%
%%%%%%%%%%%%%%%%%%%%%%%%%%%%%%%%%%%%%%%%%%%%%%%%%%%%%%%%%%%%%%%%%%
\newpage
\begin{table}
\caption{The form factors for the $B \to S$ and $B \to T$ transitions calculated
at $q^2 = m_{J/\psi}^2 ~~ ({\rm or}~~ m_K^2)$ and at the maximum momentum transfer
$q_m^2 \equiv (m_B -m_{S (T)})^2$ in the ISGW model, and
at $q^2 = m_{J/\psi}^2 ~~ ({\rm or}~~ m_K^2)$ in the ISGW2 model, respectively.
(Note that the definitions of $F_+^{B \to S}$ and ${\cal F}^{B \to T}$ are
different.  See the text.)}
\smallskip
\begin{tabular}{c|ccc}
Form factor & ISGW & ISGW(at $q_m^2$) & ISGW2
\\ \hline
   $F_+^{B \to K_0^{*+(0)}(1430)}$($q^2 = m_{J/\psi}^2; q_m^2$)
    & 0.38 & 0.61 & 0.29
\\ $F_+^{B \to f_0(1370)}$($q^2 = m_K^2; q_m^2$)
    & 0.10 & 0.68 & 0.10
\\ ${\cal F}^{B \to f_2(1270)}$($q^2 = m_K^2; q_m^2$)
    & $-0.025$ & $-0.19$ & 0.078
\end{tabular}
\end{table}
%%%%%%%%%%%%%%%%%%%%%%%%%%%%%%%%%%%%%%%%%%%%%%%%%%%%%%%%%%%%%%%%%%
%%%%%%%%%%%%%%%%%%%%%%%%%%%%%%%%%%%%%%%%%%%%%%%%%%%%%%%%%%%%%%%%%%

%%%%%%%%%%%%%%%%%%%%%%%    Table 2    %%%%%%%%%%%%%%%%%%%%%%%%%%%%
%%%%%%%%%%%%%%%%%%%%%%%%%%%%%%%%%%%%%%%%%%%%%%%%%%%%%%%%%%%%%%%%%%
\begin{table}
\caption{The branching ratios (in $10^{-6}$) of $B \to J/\psi K_0^*(1430)$
and $B \to J/\psi K_2^*(1430)$, calculated at $q^2 = m_{J/\psi}^2$ in the ISGW2
model.  For comparison, the branching ratios calculated in the original ISGW model
(using $a_2 = 0.26$) are also shown in the square bracket for the following cases:
(i) at $q^2 = m_{J/\psi}^2$ and (ii) at the maximum momentum transfer
$q_m^2 \equiv (m_B -m_{K_{0 (2)}}^*)^2$. Both cases are shown in order, such as
[(i); (ii)].  The ratios of ${\cal B}(B \to J/\psi K_0^*(1430)) /
{\cal B}(B \to J/\psi K_2^*(1430))$, which do not depend on $\xi$,
are presented as well. }
\begin{center}
\begin{tabular}{c|cccc}
Decay mode & ${\cal B}(10^{-6})[\xi=0.1]$ & ${\cal B}(10^{-6})[\xi=0.3]$
& ${\cal B}(10^{-6})[\xi=0.5]$ & ${\cal B}(10^{-6})[a_2 = 0.26]$  \\\hline
$B^+ \to J/\psi K_0^{*+}(1430)$ & 7.19 & 0.04 & 5.23 & 7.42 \\
   &  &  &  & [13.34; 34.45]  \\
$B^0 \to J/\psi K_0^{*0}(1430)$ & 6.74 & 0.04 & 4.91 & 6.96 \\
   &  &  &  & [12.51; 32.32]  \\
$B^+ \to J/\psi K_2^{*+}(1430)$ & 384.14 & 2.07 & 279.70 & 396.49 \\
   &  &  &  & [144.70; 366.87]  \\
$B^0 \to J/\psi K_2^{*0}(1430)$ & 355.56 & 1.91 & 258.90 & 366.99 \\
   &  &  &  & [133.95; 336.50]  \\\hline
${\cal B}(J/\psi K_0^{*+}(1430)) \over {\cal B}(J/\psi K_2^{*+}(1430))$
& & & & 0.019 \\
   &  &  &  & [0.092; 0.094] \\
${\cal B}(J/\psi K_0^{*0}(1430)) \over {\cal B}(J/\psi K_2^{*0}(1430))$
& & & & 0.019 \\
   &  &  &  & [0.094; 0.095]
\end{tabular}
\end{center}
\end{table}
%%%%%%%%%%%%%%%%%%%%%%%%%%%%%%%%%%%%%%%%%%%%%%%%%%%%%%%%%%%%%%%%%%
%%%%%%%%%%%%%%%%%%%%%%%%%%%%%%%%%%%%%%%%%%%%%%%%%%%%%%%%%%%%%%%%%%

%%%%%%%%%%%%%%%%%%%%%%%    Table 3    %%%%%%%%%%%%%%%%%%%%%%%%%%%%
%%%%%%%%%%%%%%%%%%%%%%%%%%%%%%%%%%%%%%%%%%%%%%%%%%%%%%%%%%%%%%%%%%
\begin{table}
\caption{The branching ratios (in $10^{-8}$) of $B \to f_0(1370) K$
and $B \to f_2(1270) K$, calculated at $q^2 = m_K^2$ in the ISGW2 model.
For comparison, the branching ratios calculated in the original ISGW model
(for $\xi =0.5$) are also shown in the square bracket for the following
cases:
(i) at $q^2 = m_K^2$ and (ii) at the maximum momentum transfer
$q_m^2 \equiv (m_B -m_{f_{0 (2)}})^2$. Both cases are shown in order, such as
[(i); (ii)].  The ratios of ${\cal B}(B \to f_0(1370) K) /
{\cal B}(B \to f_2(1270) K)$, which do not depend on $\xi$,
are presented. The CP rate asymmetries $A_{CP}$ are shown as well. }
\begin{center}
\begin{tabular}{c|ccc|c}
Decay mode & ${\cal B}(10^{-8})[\xi=0.1]$ & ${\cal B}(10^{-8})[\xi=0.3]$
& ${\cal B}(10^{-8})[\xi=0.5]$ & $A_{CP}$  \\\hline
$B^+ \to f_0(1370) K^+$ & 4.72 & 4.13 & 3.58 & 0.03  \\
   &  &  & [3.54; 157.94] & [0.03; 0.03]  \\
$B^0 \to f_0(1370) K^0$ & 0.014 & 0.041 & 0.23 & 0   \\
   &  &  & [0.22; 9.96] & [0; 0]  \\
$B^+ \to f_2(1270) K^+$ & 28.84 & 25.22 & 21.85 & 0.03  \\
   &  &  & [2.31; 124.75] & [0.004; 0.20]  \\
$B^0 \to f_2(1270) K^0$ & 0.086 & 0.25 & 1.38 & 0  \\
   &  &  & [0.15; 7.86] & [0; 0]  \\\hline
${\cal B}(f_0(1370) K^{+(0)}) \over {\cal B}(f_2(1270) K^{+(0)})$
 &  &  & 0.16 &  \\
 &  &  & [1.53; 1.27] &
\end{tabular}
\end{center}
\end{table}
%%%%%%%%%%%%%%%%%%%%%%%%%%%%%%%%%%%%%%%%%%%%%%%%%%%%%%%%%%%%%%%%%%
%%%%%%%%%%%%%%%%%%%%%%%%%%%%%%%%%%%%%%%%%%%%%%%%%%%%%%%%%%%%%%%%%%


\newpage
\begin{thebibliography}{[001]}

\bibitem{Hagiwara:fs}
K.~Hagiwara {\it et al.}  [Particle Data Group Collaboration],
%``Review Of Particle Physics,''
Phys.\ Rev.\ D {\bf 66}, 010001 (2002).
%%CITATION = PHRVA,D66,010001;%%

\bibitem{Coan:1999kh}
T.~E.~Coan {\it et al.}  [CLEO Collaboration],
%``Study of exclusive radiative B meson decays,''
Phys.\ Rev.\ Lett.\  {\bf 84}, 5283 (2000)
[arXiv:hep-ex/9912057].
%%CITATION = HEP-EX 9912057;%%

\bibitem{belle0}
Belle Collaboration, A. Ishikawa, talk at XXXVIIth Rencontres
de Moriond on Electroweak Interactions and Unified Theories, Les Arcs,
France, March 9$-$16, 2002.

\bibitem{Abe:2002av}
K.~Abe {\it et al.}  [Belle Collaboration],
%``Study of three-body charmless B decays,''
Phys.\ Rev.\ D {\bf 65}, 092005 (2002)
[arXiv:hep-ex/0201007].
%%CITATION = HEP-EX 0201007;%%

\bibitem{Kim:2002rx}
C.~S.~Kim, J.~P.~Lee and S.~Oh,
%``Nonleptonic two-body charmless B decays involving a tensor meson
%in  ISGW2 model,''
Phys.\ Rev.\ D {\bf 67}, 014002 (2003)
[arXiv:hep-ph/0205263].
%%CITATION = HEP-PH 0205263;%%

\bibitem{Kim:2001sh}
C.~S.~Kim, B.~H.~Lim and S.~Oh,
%``Charmless hadronic decays of B mesons to a pseudoscalar and
%a tensor  meson,''
Eur.\ Phys.\ J.\ C {\bf 22}, 683 (2002)
[arXiv:hep-ph/0101292].
%%CITATION = HEP-PH 0101292;%%

\bibitem{belle1}
in private communication.

\bibitem{Kim:2002rw}
C.~S.~Kim, J.~P.~Lee and S.~Oh,
%``Hadronic decays of B involving a tensor meson through a
%b $\to$ c  transition,''
Phys.\ Rev.\ D {\bf 67}, 014011 (2003)
[arXiv:hep-ph/0205262].
%%CITATION = HEP-PH 0205262;%%

\bibitem{Kim:2001py}
C.~S.~Kim, B.~H.~Lim and S.~Oh,
%``Hadronic B decays to charmless V T final states,''
Eur.\ Phys.\ J.\ C {\bf 22}, 695 (2002)
[Erratum-ibid.\ C {\bf 24}, 665 (2002)]
[arXiv:hep-ph/0108054].
%%CITATION = HEP-PH 0108054;%%

\bibitem{LopezCastro:1997im}
G.~Lopez Castro and J.~H.~Munoz,
%``Non-leptonic B decays involving tensor mesons,''
Phys.\ Rev.\ D {\bf 55}, 5581 (1997)
[arXiv:hep-ph/9702238].
%%CITATION = HEP-PH 9702238;%%

\bibitem{scalar}
S. Spanier and N. A. T\"{o}rnqvist, in \cite{Hagiwara:fs}.

\bibitem{Godfrey:1998pd}
S.~Godfrey and J.~Napolitano,
%``Light meson spectroscopy,''
Rev.\ Mod.\ Phys.\  {\bf 71}, 1411 (1999)
[arXiv:hep-ph/9811410].
%%CITATION = HEP-PH 9811410;%%

\bibitem{Close:2002zu}
F.~E.~Close and N.~A.~Tornqvist,
%``Scalar mesons above and below 1-GeV,''
J.\ Phys.\ G {\bf 28}, R249 (2002)
[arXiv:hep-ph/0204205].
%%CITATION = HEP-PH 0204205;%%

\bibitem{Cheng:2002mk}
H.~Y.~Cheng,
%``Comments on the quark content of the scalar meson f0(1370),''
Phys.\ Rev.\ D {\bf 67}, 054021 (2003)
[arXiv:hep-ph/0212361].
%%CITATION = HEP-PH 0212361;%%

\bibitem{Cheng:2002ai}
H.~Y.~Cheng,
%``Hadronic D decays involving scalar mesons,''
Phys.\ Rev.\ D {\bf 67}, 034024 (2003)
[arXiv:hep-ph/0212117].
%%CITATION = HEP-PH 0212117;%%

\bibitem{Katoch:hr}
A.~C.~Katoch and R.~C.~Verma,
%``Weak Decays Of B- And Anti-B0 Mesons To A Pseudoscalar Meson And
%A Tensor Meson Involving A B $\to$ C Transition,''
Phys.\ Rev.\ D {\bf 52} (1995) 1717
[Erratum-ibid.\ D {\bf 55} (1997) 7316].
%%CITATION = PHRVA,D52,1717;%%

\bibitem{Li:2000zb}
D.~M.~Li, H.~Yu and Q.~X.~Shen,
%``Properties of the tensor mesons f2(1270) and f2'(1525),''
J.\ Phys.\ G {\bf 27}, 807 (2001)
[arXiv:hep-ph/0010342].
%%CITATION = HEP-PH 0010342;%%

\bibitem{Ali:1997nh}
A.~Ali and C.~Greub,
%``An analysis of two-body non-leptonic B decays involving light mesons
%in the standard model,''
Phys.\ Rev.\ D {\bf 57}, 2996 (1998)
[arXiv:hep-ph/9707251].
%%CITATION = HEP-PH 9707251;%%

\bibitem{Chen:1999nx}
Y.~H.~Chen, H.~Y.~Cheng, B.~Tseng and K.~C.~Yang,
%``Charmless hadronic two-body decays of B/u and B/d mesons,''
Phys.\ Rev.\ D {\bf 60}, 094014 (1999)
[arXiv:hep-ph/9903453].
%%CITATION = HEP-PH 9903453;%%

\bibitem{Deshpande:1997ar}
N.~G.~Deshpande, B.~Dutta and S.~Oh,
%``A critical study of B decays to light pseudoscalars,''
Phys.\ Rev.\ D {\bf 57}, 5723 (1998)
[arXiv:hep-ph/9710354];
%%CITATION = HEP-PH 9710354;%%
Phys.\ Lett.\ B {\bf 473}, 141 (2000)
[arXiv:hep-ph/9712445].
%%CITATION = HEP-PH 9712445;%%

\bibitem{Diehl:2001xe}
M.~Diehl and G.~Hiller,
%``New ways to explore factorization in b decays,''
JHEP {\bf 0106}, 067 (2001)
[arXiv:hep-ph/0105194].
%%CITATION = HEP-PH 0105194;%%

\bibitem{Maltman:1999jn}
K.~Maltman,
%``The a0(980), a0(1450) and K*0(1430) scalar decay constants
%and the  isovector scalar spectrum,''
Phys.\ Lett.\ B {\bf 462}, 14 (1999)
[arXiv:hep-ph/9906267].
%%CITATION = HEP-PH 9906267;%%

\bibitem{Oh:1998wa}
S.~Oh,
%``Flavor SU(3) symmetry and factorization in B decays to two
%charmless  vector mesons,''
Phys.\ Rev.\ D {\bf 60}, 034006 (1999)
[arXiv:hep-ph/9812530].
%%CITATION = HEP-PH 9812530;%%

\bibitem{Scora:1995ty}
D.~Scora and N.~Isgur,
%``Semileptonic meson decays in the quark model: An update,''
Phys.\ Rev.\ D {\bf 52}, 2783 (1995)
[arXiv:hep-ph/9503486].
%%CITATION = HEP-PH 9503486;%%

\bibitem{Isgur:gb}
N.~Isgur, D.~Scora, B.~Grinstein and M.~B.~Wise,
%``Semileptonic B And D Decays In The Quark Model,''
Phys.\ Rev.\ D {\bf 39}, 799 (1989).
%%CITATION = PHRVA,D39,799;%%

\bibitem{Dutta:1999tg}
B.~Dutta and S.~Oh,
%``Charmless hadronic B decays and the recent CLEO data,''
Phys.\ Rev.\ D {\bf 63}, 054016 (2001)
[arXiv:hep-ph/9911263].
%%CITATION = HEP-PH 9911263;%%

\bibitem{Alam:bi}
M.~S.~Alam {\it et al.}  [CLEO Collaboration],
%``Exclusive Hadronic B Decays To Charm And Charmonium Final States,''
Phys.\ Rev.\ D {\bf 50}, 43 (1994)
[arXiv:hep-ph/9403295], and references therein.
%%CITATION = HEP-PH 9403295;%%

\end{thebibliography}
\end{document}